\documentclass[12pt]{article}

\usepackage{graphicx}

\begin{document}

%%%%%%%%%%%%%%%%%%%%%%%%%%%%%%%%%%%%%%%%%%%%%%%%%%%%%%%%%%%
\title{\bf Luneburg lens in silicon photonics}
\author{Andrea Di Falco, Susanne C Kehr, and Ulf Leonhardt$^*$\\
School of Physics and Astronomy, University of St Andrews,\\
North Haugh, St Andrews KY16 9SS, UK\\
$^*$e-mail: ulf@st-andrews.ac.uk}

\maketitle

\begin{abstract}
The Luneburg lens is an aberration-free lens that focuses light from all directions equally well. We fabricated and tested a Luneburg lens in silicon photonics. Our technique is sufficiently versatile for making perfect imaging devices on silicon platforms. 
\end{abstract}

Lenses are indispensible optical instruments, but most conventional lenses suffer from aberrations \cite{BornWolf} --- the focus depends on the direction of incidence and deteriorates off axis. The Luneburg lens \cite{Luneburg,LeoPhil} can focus light from all directions equally well. This lens establishes a classic example of non-Euclidean transformation optics \cite{LeoTyc,Huidobro,Liu,Smol1} where light in a medium is experiencing a curved spatial geometry \cite{LeoPhil}. Luneburg lenses are applied in microwave technology \cite{Cornbleet,Skolnik,Kundtz} but have remained impossible to implement in optics. Here we demonstrate an integrated Luneburg lens on a silicon chip that works for near-infrared light. Such lenses may become the building blocks of compact Fourier optics in silicon photonics \cite{Lipson}. The resolution of the Luneburg lens is limited by the wavelength of light, but our manufacturing method is also sufficiently versatile for making future perfect imaging devices \cite{Smol1,Minano,Fish} on silicon platforms.  

The Luneburg lens \cite{Luneburg,LeoPhil} is a rotationally symmetric thick lens with a spatially varying refractive-index profile that focuses light on the rim of the lens (Fig.~1). The focal point lies in the direction of the incident light; the lens thus turns the direction of a light ray into the position of the focus. In terms of light waves \cite{BornWolf}, it performs a Fourier transformation. Because of its rotational symmetry, the Luneburg lens is an ideal lens free from optical aberrations \cite{BornWolf}, but it is not a perfect lens \cite{Fish,Pendry}: its focus size is limited by the wavelength of light \cite{Colombini1} (Fig.~1). Nevertheless, it could become an important optical instrument, in particular in integrated photonics, if it were possible to make Luneburg lenses. The required index profile $n$ of the lens is currently impossible to create by doping optical materials, as the profile is given by the formula \cite{LeoPhil}
\begin{equation}
\label{lune}
n = \sqrt{2-(r/a)^2} \quad \mbox{for}\quad r \le a 
\end{equation}
where $r$ denotes the distance from the centre of the lens and $a$ its outer radius. Luneburg's formula (\ref{lune}) shows that one needs to create an index profile with a contrast of $\sqrt{2}\approx 1.4$, which is far from the reach of current doping techniques.

We have made a Luneburg lens in silicon photonics \cite{Lipson}. There light is typically confined in waveguiding structures on planar silicon chips. Instead of doping, one can exploit the three-dimensional nature of the waveguide to create an effective index profile for two-dimensional wave propagation on the chip. In our case, the Luneburg lens is a thin graded silicon disk (less than $70\mathrm{nm}$ thick) of $98\mu\mathrm{m}$ radius put between a $2\mu\mathrm{m}$ silica layer on a silicon substrate and an SU8 polymer layer on top (Fig.~2a). The total thickness of the silicon disk and the polymer is fixed to be $500\mathrm{nm}$. We exploit the fact that the effective refractive index in a waveguide depends on its geometric dimensions. As the thickness of the silicon disk is only gradually changing compared with the wavelength, we can find the local refractive index $n$ assuming a simple model: a planar sandwich of silicon substrate, silica, silicon layer, polymer and air, with each layer in the model having the local thickness of the layer in the actual device. Figure 2b shows the effective index $n$ depending on the silicon thickness for light of $1550\mathrm{nm}$ wavelength and polarized such that the magnetic field points orthogonal to the layer structure (while the electric field is parallel to the structure). One sees that the index range is sufficient to implement the Luneburg lens and other, perfectly imaging lenses such as Maxwell's fish eye \cite{Fish}. Figure 2c shows the silicon profile of our Luneburg lens (measured with a Bruker Dektak 150 stylus profiler) against the corresponding theoretical curve of formula (\ref{lune}) translated to a thickness profile by the relationship illustrated in Fig.~2b. 

Tapered waveguides like ours have been applied before for creating effective index profiles. Early Luneburg-type lenses \cite{Colombini2} were fabricated by sputtering \cite {Yao1,Yao2} but there the index contrast was insufficient for creating the proper Luneburg lens with profile (\ref{lune}). A crude Luneburg lens was made with a macroscopic, multimode waveguide \cite{Zernike}, but such a device is far too large to be integrated. Recently, approximate Eaton/Minano \cite{Minano} and Maxwell fish eye \cite{Fish} lenses were created by putting liquid droplets on a substrate \cite{Smol1}, but they cannot be integrated in a solid-state device and are not fully controllable. Luneburg lenses for surface plasmons \cite{Huidobro} were proposed and numerically studied \cite{Liu} but not made in practice. In our case, we use dielectrics that are significantly less absorptive than the metals needed for plasmons.  One could make similar index profiles with silicon metamaterials \cite{Valentine,Gabrielli,Ergin}, but here scattering losses at the structures of metamaterials become a critical limitation \cite{GabrielliFish}, whereas the index profiles made by our method are continuos and sufficiently smooth. With our technique one can controllably fabricate nearly arbitrary index profiles within the range specified in Fig.~2b on silicon photonics platforms that can combine electronics with photonics \cite{Lipson}. Such devices will work for infrared light light in the telecom band and can form the building blocks of on-chip Fourier optics.

For fabricating our Luneburg lens we used grey-scale electron-beam lithography \cite{OSKP,Reardon}. First, the prefabricated silicon on insulator chip ($2\mu\mathrm{m}$ silica, $220\mathrm{nm}$ silicon capping layer) was cleaned in acetone and IPA in an ultrasonic bath and ashed with oxygen plasma. SU8 resist of $350\mathrm{nm}$ thickness was then spun and soft-baked for $1\mathrm{min}$ at $60^o\mathrm{C}$ followed by $4\mathrm{min}$ at $100^o\mathrm{C}$. For the grey-scale lithography we used a LEO-RAITH modified electron beam with a positional accuracy of $2\mathrm{nm}$. The resist was exposed at $30\mathrm{kV}$ with a dose ranging from $0$ to $10\mu\mathrm{As/cm}^2$. Particular care had to be exercised in choosing the dose profile that produces the required silicon profile. We found empirically that a dose $D$ with profile $D_\mathrm{max}(1-r/a)^{1/2}$ results in the Luneburg profile shown in Fig.~2c. After exposure the sample was developed with EC for $45\mathrm{s}$ and hard-baked for $10\mathrm{min}$ at $180^o\mathrm{C}$. We transferred the gray-scale profile to the silicon via Reactive Ion Etching, using a blend of O2:SF6:CHF3=75:100:100 at $20\mathrm{W}$ for $90\mathrm{s}$. Finally the chip was spin coated with a layer of SU8 polymer with a maximum thickness of $500\mathrm{nm}$, soft-baked, exposed for $1\mathrm{min}$ to UV light and hard-baked to cross-link and stabilize the polymer. The sample was then cleaved for end-fire coupling (where we couple light in from the side).

After fabrication, we tested the optical performance of our Luneburg lens. The experimental measurements were performed via end-fire coupling. The beam from a C+L band amplified spontaneous emission source (centre wavelength $1575\mathrm{nm}$, bandwidth $110\mathrm{nm}$) was polarized (with the electric field parallel to the top surface of the device) and focused with a 10X objective on the facet of the sample. The beam was free to diffract in the planar waveguide until it enters the Luneburg lens. We observed the Luneburg lens from the top with a 50X Mitutoyo objective lens, with a numerical aperture of 0.55, and the image was taken by a Vidicon Electrophysics MicronViewer Camera (Spectral Response: $0.4\mu\mathrm{m}$ to $1.9\mu\mathrm{m}$). We saw a strong intensity peak at the focal spot (that originates from the focused infrared light scattered at small imperfections). To make the lens visible in our image we illuminated the device with an independent white LED. All the measurements taken with the camera were calibrated using the known radius $a$ of the lens, as measured independently with the profiler (Fig.~1c). Figure 3a shows the image acquired with the camera in false color. The light comes from the right hand side. The red spot is the focused infrared light, while the outline of the lens appears solely due the independent white LED illumination. From this figure, it is possible to measure the spot size in the two marked directions (Fig.~3b). 

The measured minimum spot size is $3.77\mu\mathrm{m}$. We expect from theory (Fig.~1) that the focal spot is about half a wavelength in size, but in our experiment it was wider. Several factors could contribute to the spot size --- the optical resolution of the microscope used in combination with the propagation in the polymer layer, the geometrical dispersion of the lens (as we used a broad band of wavelengths), scattering due to roughness and the finite width of the incident beam. The latter aspect turns out to account for the observed broadening. The input beam was not a plane wave but a Gaussian beam. The input waist was measured to be approximately $12\mu\mathrm{m}$. The Luneburg lens was a distance of $1.5\mathrm{mm}$ away from the chip facet, which means that (assuming a Gaussian beam profile) the beam reached the lens with angles between -1.1 and +1.1 degrees, with a resulting arclenght of $3.8\mu\mathrm{m}$, which agrees with the measured focus size in our Luneburg lens. 

We have performed further experiments to characterize the behavior of our Luneburg lens. In particular we launched light on the lens at different angles, to show that the lens effectively implements a spatial Fourier transform. Figure 4 shows three cases with tilting angle +4, 0 and -4 degrees in panel a), b) and c), respectively. We see that the lens focuses equally well from all directions. We also displaced the input beam with respect to the central axis of the lens, see panels d) and e) of Fig.~4. Recall that the beam size was smaller than the lens itself. Remarkably, the focused spot still appeared in the middle of the lens, confirming that the lens implements a Fourier transform, as it is mostly sensitive to the direction of propagation rather than to the spatial distribution of the beam itself. 
We have thus demonstrated a fully functional Luneburg lens in silicon photonics that can be used in integrated imaging devices. For this we exploited the three-dimensional geometric aspects of waveguides to implement refractive-index profiles for two-dimensional wave propagation that are normally impossible to make in optical materials.

\section*{Acknowledgments}
ADF is supported by an EPSRC Career Acceleration Fellowship, SCK by the University of St Andrews and UL by a Royal Society Wolfson Research Merit Award and a Blue Skies Theo Murphy Award of the Royal Society.

%%%%%%%%%%%%%%%%%%%%%%%%%%%%%%%%%%%%%%%%%%%%%%%%%%%%%%%%%%%

%%%
\begin{figure}
\begin{center}
\includegraphics[width=25.0pc]{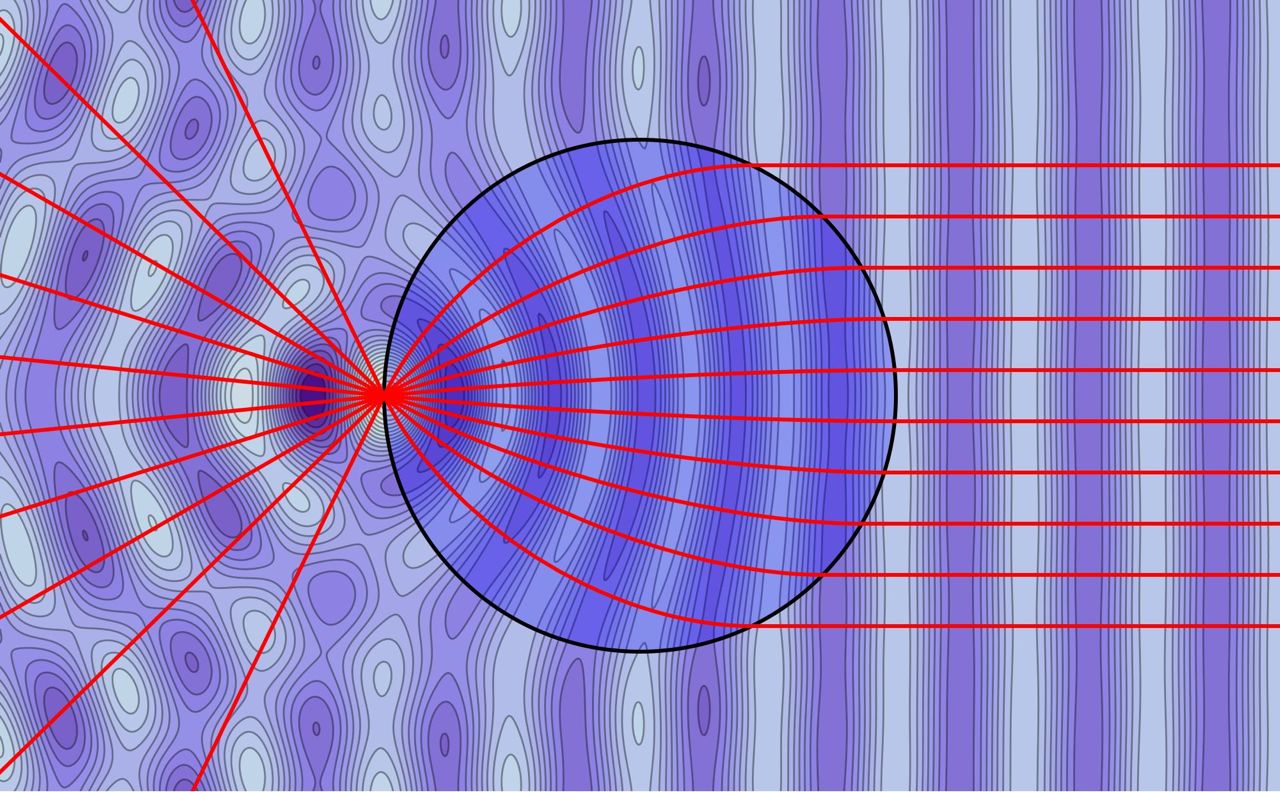}
\caption{
\small{
Luneburg lens. The lens (blue disk) focuses all light rays (red) propagating in one direction at the point on its rim that lies in that direction. The underlying wave pattern shows the real part of the Fourier component of a plane wave with wavelength $0.5a$ incident from the right (calculated by partial-wave expansion). One sees that the focal spot is about half a wavelength wide. As the lens is rotationally symmetric it focuses light from all directions equally well.}
}
\end{center}
\end{figure}
%%%

%%%
\begin{figure}
\begin{center}
\includegraphics[width=20.0pc]{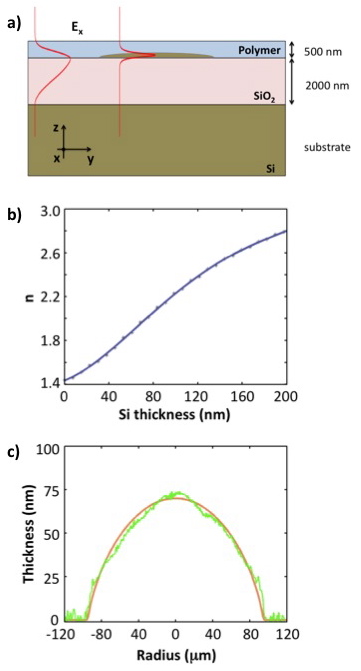}
\caption{
\small{
The device. {\bf a}:  the waveguide confining the light also creates the refractive index profile (\ref{lune}) of the Luneburg lens for horizontal light propagation (Fig.~1) on a chip. The red curves show the vertical intensity distributions. {\bf b}: effective refractive index depending on the thickness of the silicon below the polymer in (a). {\bf c}: measured silicon profile (green) versus the theoretical curve (red) required for implementing the Luneburg lens.}
}
\end{center}
\end{figure}
%%%

%%%
\begin{figure}
\begin{center}
\includegraphics[width=35.0pc]{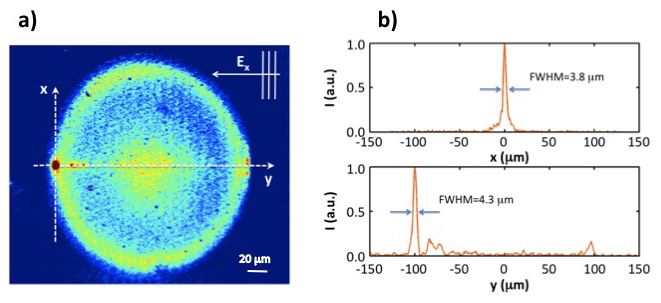}
\caption{
\small{
Light focusing. {\bf a}: False-colour image of the observed intensity profile in the Luneburg lens (Fig.~2). Infrared light incident from the right is guided on the chip and focused in the lens (red spot). The lens itself is made visible by illuminating it with white light from the top. {\bf b}: measured intensity profiles of the focused infrared light along the two lines indicated in (a). In our experiment the full width at half maximum (FWHM) of the focus is dominated by the width of the incident Gaussian beam.}
}
\end{center}
\end{figure}
%%%

%%%
\begin{figure}
\begin{center}
\includegraphics[width=32.0pc]{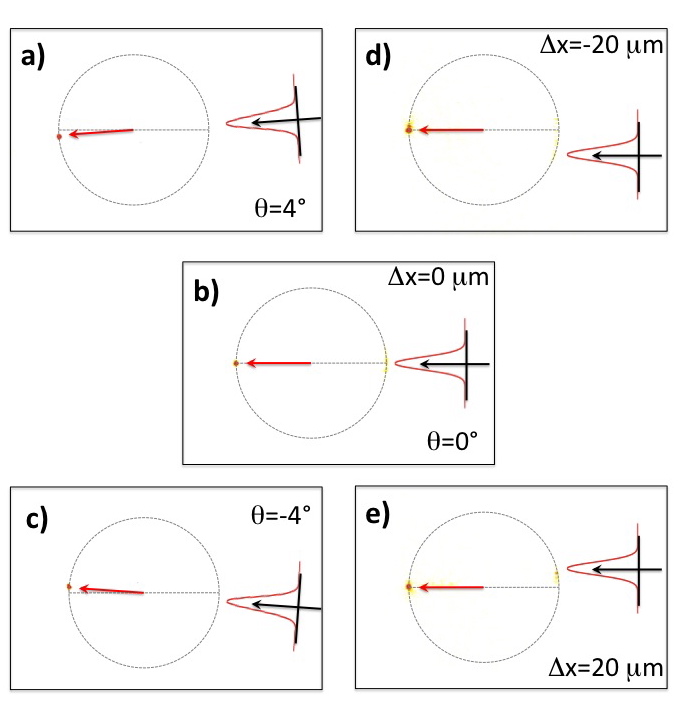}
\caption{
\small{
Performance tests. The figure shows the observed infrared light focused in the Luneburg lens (yellow-red spots similar to Fig.~3, except that the lens is not made visible) depending on the propagation direction (red arrow with angle $\theta$) and the offset $\Delta x$ of the incident Gaussian beam (profile shown). {\bf a}-{\bf c}: the focal spot lies in the propagation direction. {\bf d},{\bf e}: the focal point is independent of the offset. Our device thus behaves as expected from a Luneburg lens.}
}
\end{center}
\end{figure}
%%%

\end{document}